\documentclass[aps,prd,preprint]{revtex4}
\usepackage{mathrsfs}
\usepackage{epsfig}
 \draft
\begin{document}
\author{Yu Jiang$^{1}$, Liu-Jun Luo$^{2}$,
and Hong-Shi Zong$^{2,3}$}
\address{$^{1}$ College of Mathematics, Physics and Information
Engineering, Zhejiang Normal University, 688 Yingbin Avenue, Jinhua
City, Zhejiang Province 321004, China}
\address{$^{2}$ Department of Physics, Nanjing University,
Nanjing 210093, China}
\address{$^{3}$Joint Center for Particle,
Nuclear Physics and Cosmology, Nanjing 210093, China}

\title{A model study of quark number susceptibility at
finite temperature beyond rainbow-ladder approximation}

\begin{abstract}
In this paper we calculate the quark number susceptibility (QNS) of
QCD at finite temperature under the rainbow-ladder and Ball-Chiu
type truncation schemes of the Dyson-Schwinger approach. It is found
that the difference between the result of the rainbow-ladder
truncation and that of Ball-Chiu type truncation is small, which
shows that the dressing effect of the quark-gluon vertex on the QNS
at finite temperature is small. It is also found that at low
temperature the quark number susceptibility is nearly zero and it
increases sharply when the temperature approaches the chiral phase
transition point. A comparison between the result in the present
paper with those in the literature is made.


E-mail: zonghs@chenwang.nju.edu.cn
\end{abstract}

\maketitle

\section{Introduction}

As is well known, dynamical chiral symmetry breaking (DCSB) and
confinement are two fundamental features of quantum chromodynamics
(QCD). Nowadays it is generally believed that at high enough
temperature and/or density strongly interacting matter will undergo
a chiral symmetry restoring and deconfining phase transition (for a
recent review, see for example Ref. \cite{r1}). The study of such phase
transitions has become a field of great theoretical and experimental
interest over past years because it will provide us a fundamental
understanding of the basic theory including the origin of observed
mass and the nature of the early universe. A central goal of
relativistic heavy ion collision experiments is to explore the phase
structure and investigate the chiral phase transition of QCD matter.

In the study of chiral phase transition the enhanced quark number
fluctuations are thought to be an essential characteristic
\cite{a1,a2,a3,a4,a5,a6}. In the confined/chirally broken phase
quark (baryon) numbers are associated with hadrons in integer units
whereas in the deconfined/chirally restored phase they are
associated with quarks in fractional units which could lead to
different quark number fluctuations in the two phases. Theoretically
such fluctuation is constructed from measurement of quark number
susceptibility (QNS) as a function of temperature, which is the the
response of quark number density to infinitesimal change of chemical
potential at zero chemical potential \cite{a1}. In particular, it
was recently argued that the QNS may be used to identify the chiral
critical point in the QCD phase diagram \cite{a7,a8,a9}. Hence the
QNS of QCD has been extensively studied over the past twenty years
using different approaches including lattice QCD simulations
\cite{a3,L1,L2,L3,L4,L5,L6}, Nambu-Jona-Lasinio model \cite{a4,NJL},
hard-thermal/dense-loop resummation techniques
\cite{HTL1,HTL2,HTL3,HTL4,HTL5} and the rainbow-ladder approximation of the Dyson-Schwinger equations (DSEs) approach \cite{He1,He2,He3}. It should be noted that many model calculations of the QNS are done at the level of mean field approximation (it corresponds to the rainbow-ladder approximation in the framework of the DSEs approach) \cite{He1,He2,He3}.
Since the QNS is a very important quantity in characterizing the chiral phase transition,
theoretically it is very interesting to calculate this quantity beyond mean field approximation, i.e, beyond the rainbow-ladder approximation of the DSEs approach. The purpose of the present work is to calculate the QNS at finite temperature beyond the rainbow-ladder approximation of the DSEs approach.

Over the past few years considerable progress has been made in the
framework of the rainbow-ladder approximation of the DSEs
approach \cite{DSE1,DSE2,DSE3,DSE4,DSE5,DSE6} (for a recent
review, see Ref. \cite{DSE7}). Due to the great success of the
rainbow-ladder approximation (RL) of the DSEs approach in hadron physics,
the authors in Refs. \cite{He1,He2} adopt this approximation to solve DSEs
for the dressed quark propagator and the vector vertex and from
these obtain the numerical value of the QNS. However, it is well
known that the rainbow-ladder approximation uses a bare quark-gluon
vertex, which violates Slavnov-Taylor identity of QCD. In order to
overcome this deficiency, physicists are trying their best to go
beyond the rainbow-ladder approximation. Much work has been done in this direction,
including the Ball-Chiu (BC) vertex derived from the vector
Ward-Takahashi identity (WTI) \cite{BC1,BC2}, the CP vertex
\cite{CP} which takes into account some transverse effects and the
vertex derived from the transverse WTI \cite{TWTI1,TWTI2,TWTI3}. As
was shown in Ref. \cite{QCD1}, if the ghost amplitudes and the gluon
dressing function factor are ignored in Slavnov-Taylor identity one
would find the resultant relation has the form of a color matrix times
the WTI structure which could be satisfied by the BC vertex ansatz
multiplied by the color matrix. Therefore, in this paper we adopt
such a quark-gluon vertex ansatz to explore the effect of vertex
dressing on the quark number susceptibility.

When the BC vertex is used to calculate the dressed quark propagator
from the DSEs, one should construct a consistent kernel approximation
corresponding to this vertex. How to construct systematic and
convergent expansions for the kernels of DSEs is a long-standing
unsolved problem. Recently, an important progress in this aspect has
been achieved in Ref. \cite{DSE8}. The authors in Ref. \cite{DSE8} have
proposed a Bethe-Salpeter kernel which is valid for a general
quark-gluon vertex and this provides a theoretical foundation for
calculating the QNS beyond the rainbow-ladder approximation. In the
present paper we will use this method to calculate the QNS at finite
temperature.

\section{General formula for QNS}

To make this paper self-contained, let us first recall the
definition of the QNS which is analogous to the familiar magnetic
susceptibility. If we confine ourselves to the two-flavor case with
exact isospin symmetry and set $\mu_u=\mu_d=\mu$ ($\mu_u$ and
$\mu_d$ are the chemical potential of the up and down quarks), the
quark number density and the corresponding susceptibility are given
by
\begin{equation}
\rho(\mu,T)=\frac{T}{V}\frac{\partial \ln Z(\mu,T)}{\partial \mu},
\label{rho0}
\end{equation}
and
\begin{equation}
\chi(T)=\left.\frac{\partial \rho(\mu,T)}{\partial
\mu}\right|_{\mu=0},\label{chi1}
\end{equation}
respectively, where $Z(\mu,T)$ is the partition function of QCD at
finite temperature $T$ and quark chemical potential $\mu$.

From Eq. (\ref{rho0}) and by using functional integral techniques, one
can derive a well-known result (see, for example, Refs. \cite{zong1,zong2})
\begin{eqnarray}
\rho(\mu,T)&=&-N_cN_fT\sum\limits_{k=-\infty}^{+
\infty}\int\frac{d^3\vec{p}}{(2\pi)^3}\mbox{tr}\left\{G[\mu](p_k)
\gamma_4\right\},\label{rho1}
\end{eqnarray}
where $G[\mu](p_k)$ is the dressed quark propagator at finite $T$
and $\mu$, $p_k=(\vec p,\omega_k)$ and $\omega_k=(2k+1)\pi T$ are
the fermion Matsubara frequencies with $k$ being integers, $N_c$ and $N_f$
denote the number of colors and of flavors, respectively, and the
trace operation is over Dirac indice. From Eq. (\ref{rho1}) it can be
seen that the quark number density $\rho(\mu,T)$ is totally
determined by the dressed quark propagator at finite chemical
potential and temperature. If one substitutes the free quark
propagator at finite $T$ and $\mu$ into Eq. (\ref{rho1}), one will
obtain exactly the Fermi statistics result for the quark number
density of a free quark gas, as was shown in Ref. \cite{zong2}.

Substituting Eq. (\ref{rho1}) into Eq. (\ref{chi1}), one finds the
following expression
\begin{eqnarray}
\chi(T)&=&-N_cN_fT\sum_k\int\frac{d^3\vec{p}}{(2\pi)^3}\mbox{tr}
\left\{\frac{\partial
G[\mu](p_k)}{\partial\mu}\gamma_4\right\}_{\mu=0}.
\end{eqnarray}
By means of the following identity
\begin{eqnarray}
\left.\frac{\partial G^{-1}[\mu](p_k)}
{\partial\mu}\right|_{\mu=0}&=&-\Gamma_4(p_k,0),
\end{eqnarray}
where $\Gamma_4(p_k,0)$ is the fourth component of the dressed
vector vertex $\Gamma_\mu(p_k,0)$ (the chemical potential $\mu$
can be regarded as the fourth component of a constant external
vector field $\mathcal{V}_\mu$ which is coupled to the vector current
of quarks, see Ref. \cite{He1}), one can find the following general
formula for the QNS at finite $T$
\begin{eqnarray}
\chi(T)&=&-N_cN_fT\sum\limits_{k=-\infty}^{+
\infty}\int\frac{d^3\vec{p}}{(2\pi)^3}\mbox{tr}[G(\vec{p},
\omega_k)\Gamma_4(p_k,0)G(\vec{p},\omega_k)\gamma_4],\label{chi2}
\end{eqnarray}
which is the same as the expression given in Ref. \cite{He1}. It should
be noted that in contrast to the chiral susceptibility \cite{He4},
there is no ultraviolet divergence in the quark number
susceptibility, and if one substitutes the free quark propagator (in
the chiral limit) at finite $T$ and the bare vertex into Eq. (\ref{chi2}), one will obtain exactly the QNS of a free massless
quark gas $\chi_{free}=N_fT^2$ \cite{He1}.

Since Eq. (\ref{chi2}) is a model-independent expression for the QNS at finite $T$, one can
obtain the exact result of the QNS at finite $T$ once the exact dressed
quark propagator and vector vertex at finite $T$ is known. However, at
present it is very difficult to calculate the dressed quark
propagator and vector vertex from first principles of QCD and hence
one has to resort to various nonperturbative QCD models. In the
following we will use DSEs of QCD to calculate the QNS from Eq. (\ref{chi2}).

\section{DSEs approach}

At finite temperature $T$ the Dyson-Schwinger equation for the quark
propagator can be written as follows (the renormalization
constants are set to one and this will be explained later)
\begin{eqnarray}
G^{-1}(p_k)&=&G^{-1}_0(p_k)+\frac{4}{3}T\sum_n\int\frac{d^3\vec
q}{(2\pi)^3}g^2D_{\mu\nu}(p_k-q_n)\gamma_\mu
G(q_n)\Gamma_\nu^g(q_n,p_k),\label{quarkDSE1}
\end{eqnarray}
where $G^{-1}_0$ is the inverse of the free quark propagator, $g$ is
the strong coupling constant, $D_{\mu\nu}$ is the dressed gluon
propagator, $\Gamma_\nu^g$ is the dressed quark-gluon vertex,
$q_n=(\vec q,\omega_n)$ and $\omega_n=(2n+1)\pi T$ with $n$ being
integers. At finite temperature $T$ the $O(4)$ symmetry of
QCD is broken to $O(3)$ (in the present paper we shall always work
in Euclidean space). From general Lorentz structure analysis one
finds that the inverse of the dressed quark propagator at finite $T$
can be decomposed as follows \cite{DSE9}
\begin{eqnarray}
G^{-1}(p_k)&=&i\vec\gamma\cdot\vec pA(\vec
p^2,\omega_k^2)+i\omega_k\gamma_4C(\vec p^2,\omega_k^2)+B(\vec
p^2,\omega_k^2),\label{quarkdcmp1}
\end{eqnarray}
where $A$, $B$ and $C$ are scalar functions of $\vec p^2$ and
$\omega_k^2$.

The dressed vector vertex $\Gamma_\mu$ satisfy the following
inhomogeneous Bethe-Salpeter equation
\begin{eqnarray}
\Gamma_\mu(p_k,P_{\Omega_l})&=&\gamma_\mu +T\sum_n\int\frac{d^3\vec
q}{(2\pi)^3}\big[G(q_+)\Gamma_\mu(q_n,P_{\Omega_l})G(q_-)\big]_{sr}
K^{rs}_{tu}(q_n,p_k;P_{\Omega_l}),\label{vertexDSE1}
\end{eqnarray}
where $p_k$ is the relative and $P_{\Omega_l}$ the total momentum of
quark-antiquark pair, $P_{\Omega_l}=(\vec P, \Omega_l)$ and
$\Omega_l=2l\pi T$ with $l$ being integers, $q_\pm=q_n\pm P_{\Omega_l}/2$
and $r,s,t,u$ represent color and Dirac indices. Here $K$ is the
fully-amputated quark-antiquark scattering kernel. According to
Eq. (\ref{chi2}), when calculating the QNS we need to know the fourth
component of the dressed vector vertex at zero total momentum
$\Gamma_4(p_k,0)$ which can be decomposed as follows (see the appendix):
\begin{eqnarray}
\Gamma_4(p_k,0)&=&\omega_k(F_1\vec\gamma\cdot\vec
p+F_2\omega_k\gamma_4-iF_3),\label{verdcmp1}
\end{eqnarray}
where $F_1$, $F_2$ and $F_3$ are scalar functions of $\vec p^2$ and
$\omega_k^2$. For the free field case, $F_1=F_3=0$ and
$F_2=1/\omega_k^2$. Substituting Eqs. (\ref{quarkdcmp1}) and
(\ref{verdcmp1}) into Eq. (\ref{chi2}), one finds the following
expression for the QNS at finite temperature
\begin{eqnarray}
\chi(T)&=&\frac{2N_cN_fT}{\pi^2}\sum_k\int_0^\infty dp\,
p^2\omega_k^2\frac{2BCF_3+2ACF_1\vec p^2+F_2(C^2\omega_k^2-A^2\vec
p^2-B^2)}{(A^2\vec p^2+C^2\omega_k^2+B^2)^2},\label{chi3}
\end{eqnarray}
and if one uses the free field value $A=C=1$, $B=0$ and $F_1=F_3=0$,
$F_2=1/\omega_k^2$ in Eq. (\ref{chi3}) one will re-obtain the free
quark gas result.

When one tries to solve Eqs. (\ref{quarkDSE1}) and
(\ref{vertexDSE1}), one needs to input the model gluon
propagator in advance. In the present work we employ the
following model gluon propagator at finite temperature
($Q_{\Omega_l}=(\vec Q, \Omega_l)$):
\begin{eqnarray}
g^2D_{\mu\nu}(Q_{\Omega_l})&=&P_{\mu\nu}^T(Q_{\Omega_l})D_T
+P_{\mu\nu}^L(Q_{\Omega_l}) D_L,
\end{eqnarray}
where $P_{\mu\nu}^{T,L}$ are the transverse and longitudinal projection
operators, respectively:
\begin{eqnarray}
P_{\mu\nu}&=&\delta_{\mu\nu}-\frac{Q_\mu Q_\nu}{Q_{\Omega_l}^2},\\
P^T_{\mu\nu}&=&\left\{\begin{array} {l@{\quad{}\quad}l}0&\mu\,\,
\mathrm{and/or}\,\, \nu=4\\{}\delta_{ij}-Q_iQ_j/\vec
Q^2&\mu,\nu=i,j=1,2,3\end{array}\right.,\\
P_{\mu\nu}^L&=&P_{\mu\nu}-P_{\mu\nu}^T,
\end{eqnarray}
and
\begin{eqnarray}
D_T&=&\frac{4\pi^2D}{\omega^6}Q_{\Omega_l}^2
e^{-Q_{\Omega_l}^2/\omega^2},\label{gluonT}\\
D_L&=&\frac{4\pi^2D}{\omega^6}(Q_{\Omega_l}^2+m_g^2)
e^{-(Q_{\Omega_l}^2+m_g^2)/\omega^2}.\label{gluonL}
\end{eqnarray}
This model gluon propagator is a simplified form of the effective
interaction proposed in Ref. \cite{DSE10}, which is the finite
temperature extension of Maris-Tandy model \cite{DSE1,DSE2}. It
should be noted that this model delivers an ultraviolet finite gap
equation for the quark propagator and hence the regularization
mass-scale could be removed to infinity and the renormalization
constants set to one. In this model $m_g^2=16\pi^2T^2/5$ is a
temperature-dependent mass-scale and $\omega$ and $D$ are active
parameters of the model which are not independent: a change in $D$
will be compensated by an alternation of $\omega$ \cite{DSE4}. For $\omega\in[0.3,0.5]$ GeV
in the rainbow-ladder truncation scheme, fitted in-vacuum low-energy observables are approximately constant if $\omega D=(0.8\mbox{GeV})^3$ and in the present paper we use
$\omega=0.5$ GeV.

Now let us discuss the truncation schemes of DSEs. Under the rainbow
approximation one uses bare vertex $\gamma_\nu$ to replace
$\Gamma^g_\nu$ and Eq. (\ref{quarkDSE1}) becomes (in the chiral limit)
\begin{eqnarray}
G^{-1}(p_k)&=&i\vec\gamma\cdot\vec
p+i\omega_k\gamma_4+\frac{4}{3}T\sum_n\int\frac{d^3\vec
q}{(2\pi)^3}g^2D_{\mu\nu}(p_k-q_n)\gamma_\mu
G(q_n)\gamma_\nu.\label{quarkDSE2}
\end{eqnarray}
Under the corresponding ladder approximation for the fully-amputated
quark-antiquark scattering kernel $K$ one uses
\begin{eqnarray}
K^{rs}_{tu}(q_n,p_k;P_{\Omega_l})&=&-g^2D_{\mu\nu}(q_n-p_k)
\left(\gamma_\mu\frac{\lambda^a}{2}\right)_{tr}
\left(\gamma_\nu\frac{\lambda^a}{2}\right)_{su},
\end{eqnarray}
and Eq. (\ref{vertexDSE1}) becomes
\begin{eqnarray}
\Gamma_\mu(p_k,P_{\Omega_l})&=&\gamma_\mu
-\frac{4}{3}T\sum_n\int\frac{d^3\vec
q}{(2\pi)^3}g^2D_{\rho\sigma}(p_k-q_n)\gamma_\rho
G(q_+)\Gamma_\mu(q_n,P_{\Omega_l})G(q_-)\gamma_\sigma.\label{vertexDSE2}
\end{eqnarray}
The rainbow-ladder approximation is the lowest order truncation scheme
for the DSEs \cite{DSE11}, and due to the reasons mentioned in
the introduction physicists are trying to go beyond it for years.
Here the key points are the dressed quark-gluon vertex and the
corresponding quark-antiquark scattering kernel. Just as is shown in the introduction, in this work, for the dressed
quark-gluon vertex we will employ the following finite temperature
extension \cite{DSE12} of BC vertex \cite{BC1,BC2}
\begin{eqnarray}
\Gamma^{BC}_\mu&=&(\vec\Gamma^{BC},\Gamma_4^{BC}),\label{BCVer1}
\end{eqnarray}
where
\begin{eqnarray}
\vec\Gamma^{BC}(q_n,p_k)&=&\Sigma_A\,\vec\gamma+(\vec q+\vec
p)\left[\frac{\vec \gamma\cdot(\vec q+\vec
p)}{2}\Delta_A+\frac{\gamma_4(\omega_n+\omega_k)}{2}\Delta_C-i\Delta_B\right],\\
\Gamma_4^{BC}(q_n,p_k)&=&\Sigma_C\,\gamma_4+
(\omega_n+\omega_k)\left[\frac{\vec \gamma\cdot(\vec q+\vec
p)}{2}\Delta_A+\frac{\gamma_4(\omega_n+\omega_k)}{2}\Delta_C-i\Delta_B\right],
\end{eqnarray}
with ($\mathcal{F}=A,B,C$)
\begin{eqnarray}
\Sigma_\mathcal{F}(\vec q^2,\omega_n^2,\vec p^2,\omega_k^2)&=&
\frac{\mathcal{F}(\vec q^2,\omega_n^2)+\mathcal{F}(\vec
p^2,\omega_k^2)}{2},\\
\Delta_\mathcal{F}(\vec q^2,\omega_n^2,\vec p^2,\omega_k^2)&=&
\frac{\mathcal{F}(\vec q^2,\omega_n^2)-\mathcal{F}(\vec
p^2,\omega_k^2)}{q_n^2-p_k^2}.
\end{eqnarray}
Therefore Eq. (\ref{quarkDSE1}) becomes
\begin{eqnarray}
G^{-1}(p_k)&=&i\vec\gamma\cdot\vec
p+i\omega_k\gamma_4+\frac{4}{3}T\sum_n\int\frac{d^3\vec
q}{(2\pi)^3}g^2D_{\mu\nu}(p_k-q_n)\gamma_\mu
G(q_n)\Gamma_\nu^{BC}(q_n,p_k).\label{quarkDSE3}
\end{eqnarray}

Now one should find a kernel for Eq. (\ref{vertexDSE1}) which is
consistent with BC vertex. This is a difficult task and recently
great progress has been done on this aspect. The authors in Ref. \cite{DSE8} find a way to constrain the kernel for a general vertex. For BC vertex, following their method, Eq. (\ref{vertexDSE1}) can be written as
\begin{eqnarray}
\Gamma_\mu(p_k,0)&=&\gamma_\mu-\frac{4T}{3}\sum_n\int\frac{d^3\vec
q}{(2\pi)^3}g^2D_{\rho\sigma}(p_k-q_n)\gamma_\rho
G(q_n)\Gamma_\mu(q_n,0)G(q_n)\Gamma^{BC}_\sigma+\nonumber\\
&&{}+\frac{4T}{3}\sum_n\int\frac{d^3\vec
q}{(2\pi)^3}g^2D_{\rho\sigma}(p_k-q_n)\gamma_\rho
G(q_n)\Lambda_{\mu\sigma}(p_k,q_n;0),\label{vertexDSE3}
\end{eqnarray}
where $\Lambda_{\mu\sigma}(p_k,q_n;0)$ is a four-point Schwinger
function which is completely defined by the quark self-energy
\cite{DSE13,DSE14}. It satisfies a similar identity as those in Ref. \cite{DSE8}
\begin{eqnarray}
i(p_k-q_n)_\sigma\Lambda_{\mu\sigma}(p_k,q_n;0)&=&\Gamma_\mu(p_k,0)
-\Gamma_\mu(q_n,0),\label{vWTI}
\end{eqnarray}
and from this identity one can write down the expressions of
$\Lambda_{\mu\sigma}(p_k,q_n;0)$ (for details, see the appendix). Now, by means of the BC vertex ansatz and the corresponding quark-antiquark scattering kernel in Eq. (\ref{vertexDSE3}) (in this
paper we shall call this truncation scheme a BC-type truncation), we can numerically calculate the dressed quark propagator and the vector vertex at finite $T$ beyond the rainbow-ladder approximation, which is needed to calculate the QNS of QCD at finite temperature.

\section{Numerical results}
By using numerical iteration method one can solve Eqs. (\ref{quarkDSE2}), (\ref{vertexDSE2}),
(\ref{quarkDSE3}) and (\ref{vertexDSE3}), and
the results are shown in Figs 1-4. Here it should be noted that when working with the
rainbow-ladder truncation, we employ model parameters in which
$D=1.0$ GeV$^2$. Now when calculating the dressed quark propagator using the BC-type truncation, we employ the model gluon propagator of the same form. However, because the amount of chiral symmetry breaking (as measured by the chiral condensate) and related quantities such as the
pion decay constant are very different between the rainbow-ladder truncation and the BC-type truncation \cite{DSE8}, one should employ refitted model parameters in Eqs. (\ref{gluonT}) and (\ref{gluonL}) \cite{shi1,shi2} when performing the calculation in the BC-type truncation. Under the BC vertex, the value of the parameter $D$ fitted from the chiral condensate is $D=0.5 ~\mathrm{GeV}^2$ (see, Refs. \cite{shi1,shi2}), and in this paper we choose this value.

The dressing functions $A(0,\pi^2 T^2)$, $C(0,\pi^2 T^2)$ and $B(0,\pi^2 T^2)$ as
functions of temperature are shown in Fig. 1 and Fig. 2, respectively. If one takes $B(0,\pi^2
T^2)$ as the chiral order parameter, from Fig. 2 it can be seen that the transition temperature $T_c\sim102$ MeV for both RL and BC-type truncations, because at this temperature $B(0,\pi^2
T^2)$ for both two cases decrease to zero. This temperature is smaller
than the result of lattice QCD in which $T_c\sim 150$ MeV (see, for
example, Ref. \cite{Lat1}) and this discrepancy may be ascribed to the fact that we have ignored the perturbative tail in our model gluon propagator (\ref{gluonT}) and (\ref{gluonL}). From Fig. 1 it can be seen that the
derivative of $A(0,\pi^2 T^2)$ and $C(0,\pi^2 T^2)$ with respect to $T$ undergo
a sudden change at $T_c$. The temperature dependence of $F_1(0,\pi^2
T^2)$, $F_2(0,\pi^2 T^2)$ and $F_3(0,\pi^2 T^2)$ is shown in
Figs. 3-4. One can see when $T>T_c$, the main dressing effect of the
vector vertex $\Gamma_4(p_k,0)$ comes from $F_2$. This is
reasonable because from Eq. (\ref{verdcmp1}) $F_2\omega_k^2$ is the
coefficient of the bare vertex $\gamma_4$, which should be a dominant
term when $T$ is high enough.

The calculated QNS under both truncation schemes are shown in Fig.
5 and Fig. 6. In Fig. 5 $\chi_{BC}$ and $\chi_{RL}$ are the QNS obtained using
BC-type truncation and RL truncation, respectively. $\langle\bar
q q\rangle_{BC}$ and $\langle\bar q q\rangle_{BC0}$ are the chiral
condensate obtained using BC-type truncation at finite T and zero T, respectively.
$\langle\bar q q\rangle_{RL}$ and $\langle\bar q q\rangle_{RL0}$ are
the chiral condensate obtained using RL truncation at finite T and zero T,
respectively. It can be seen from Fig. 5 that when $T<T_c$, the
QNS increases with increasing $T$ and the rate of increasing also
increases with $T$. At $T=T_c$, $\chi/\chi_{free}\sim0.95$ and the
derivative of QNS with respect to $T$ undergoes a sudden change (from positive value to
negative value). Such a behavior could be regarded as the signal of chiral phase
transition. When $T_c<T\lesssim200$ MeV, the QNS decreases slowly
with increasing $T$. From Fig. 6 it can be seen that when $T$ is larger
than about $200$ MeV, the QNS again increases very slowly towards the
free quark gas limit with increasing $T$, which is the result of
asymptotic freedom of QCD. In all ranges of the temperature investigated,
the QNS obtained using BC-type truncation is slightly larger than the one obtained using RL
truncation, but the difference is smaller than $10\%$ when $T>95$ MeV,
which indicates that RL approximation is good for calculating the
QNS at finite $T$. The chiral condensates obtained using both truncation schemes
are also shown in Fig. 5, in which they are normalized,
respectively, by the values of the zero temperature results obtained using the
corresponding truncation schemes. From Fig. 5 it can be seen that when $T<T_c$ the
chiral condensate decreases monotonously with increasing temperature, and when $T\geq T_c$ it equals zero. One can also see that the chiral condensates obtained using the two truncation
schemes are almost the same, which shows that the dressing effect of the quark-gluon vertex on the chiral condensate is very small.

\begin{figure}[t]
\begin{minipage}[t]{7cm}
  \includegraphics[width=6.2cm]{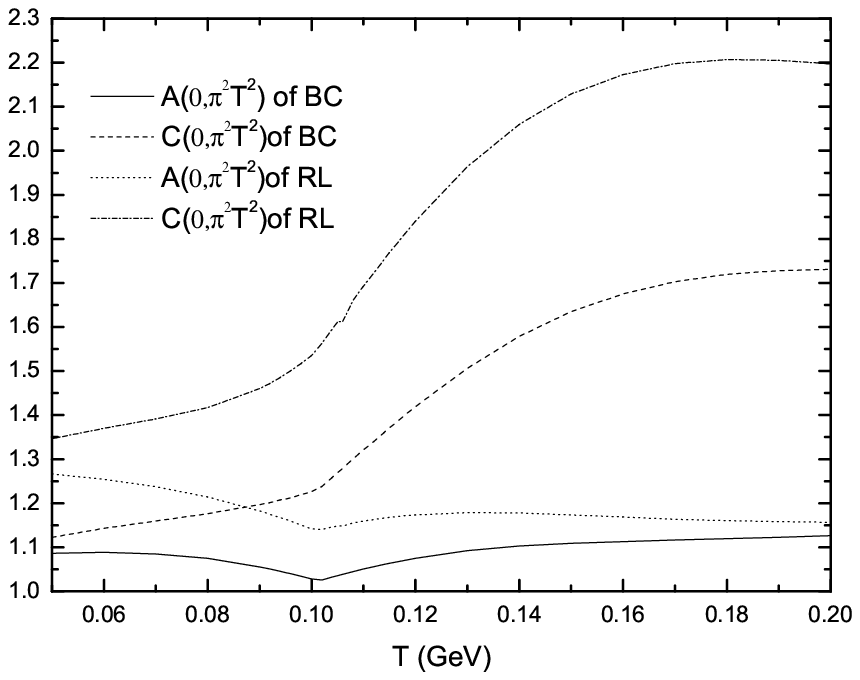}\\
  \parbox{7cm}{\caption{ $A(0,\pi^2T^2)$ and
  $C(0,\pi^2T^2)$ of the quark propagator.}}
\end{minipage}
\hfill
\begin{minipage}[t]{7cm}
  \includegraphics[width=6.4cm]{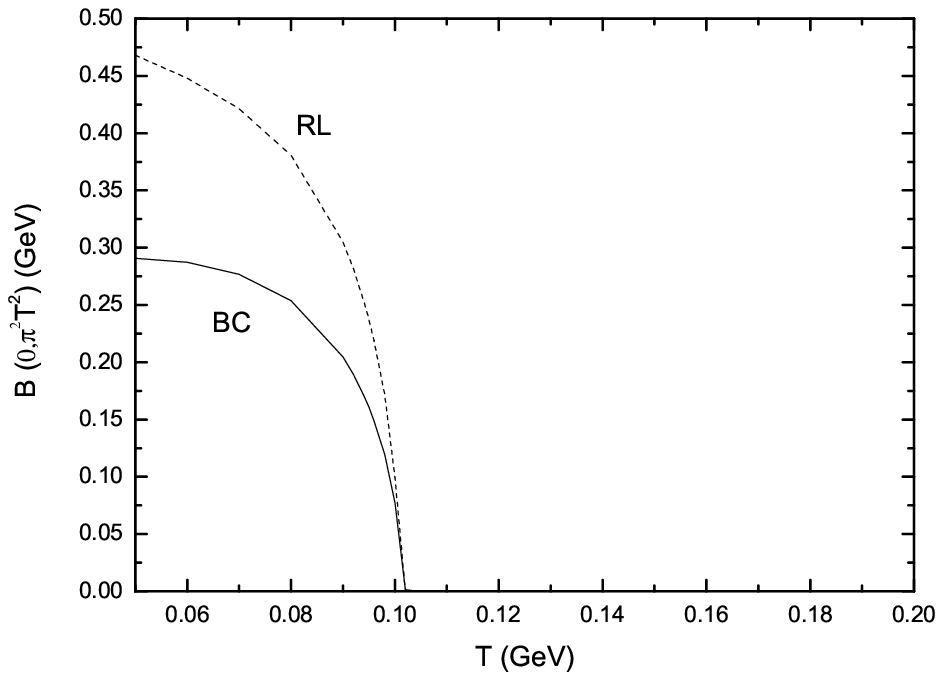}\\
  \parbox{7cm}{\caption{$B(0,\pi^2T^2)$ of the quark propagator.}}
  \end{minipage}
\end{figure}
\begin{figure}[t]
\begin{minipage}[t]{7cm}
  \includegraphics[width=6cm]{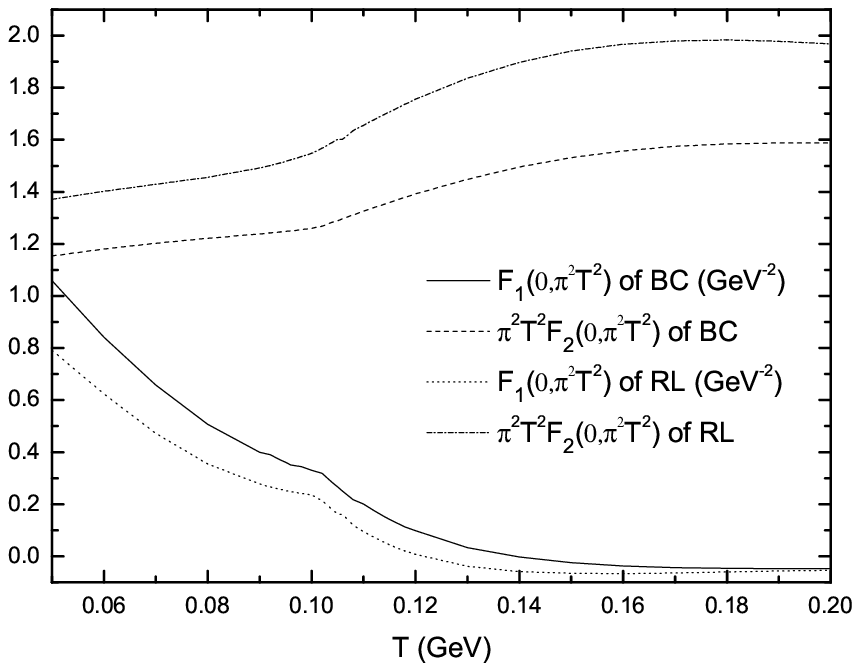}\\
  \parbox{7cm}{\caption{ $F_1(0,\pi^2T^2)$ and
  $F_2(0,\pi^2T^2)$ of the vector vertex.}}
\end{minipage}
\hfill
\begin{minipage}[t]{7cm}
  \includegraphics[width=6.4cm]{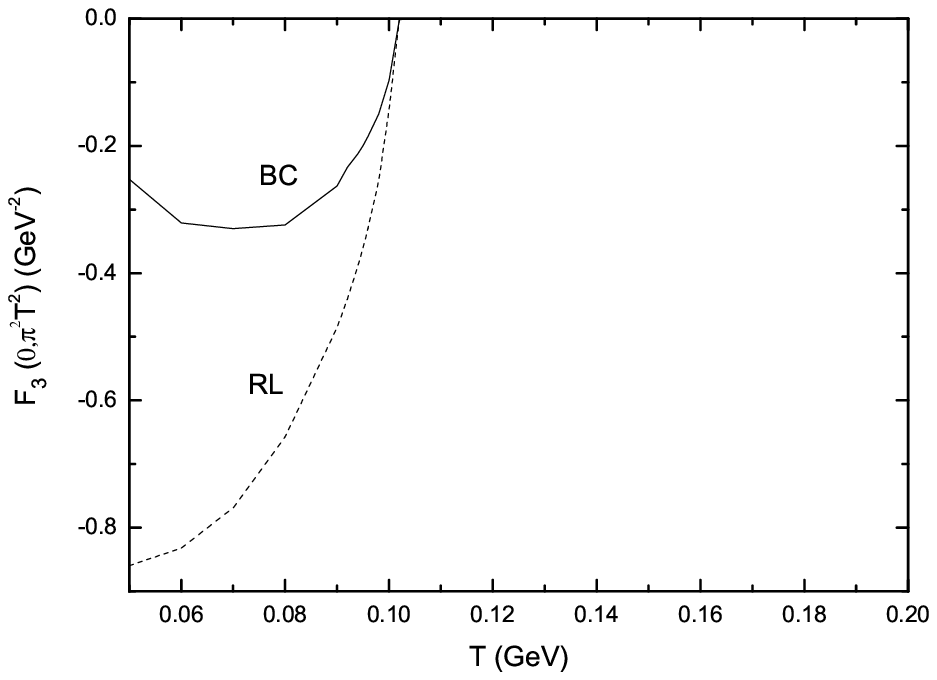}\\
  \parbox{7cm}{\caption{$F_3(0,\pi^2T^2)$ of the vector vertex.}}
  \end{minipage}
\end{figure}

It is instructive to compare our result of QNS with the corresponding results given in Refs. \cite{He2} and \cite{He3}. In Ref. \cite{He3} a parametrized quark
propagator which is based on an effective interaction similar with
the one used in the present paper is employed and the resultant QNS
exhibits similar behaviors with the result of the present paper.
However, the QNS given in Ref. \cite{He3} becomes negative when
$50~\mbox{MeV}<T<90~\mbox{MeV}$, and according to the result of the
present paper, such a behavior might be an artifact of the model quark
propagator used in Ref. \cite{He3}. On the other hand, in Ref. \cite{He2} a
rank-1 separable model for the effective gluon propagator is employed
and the resultant QNS has a peak at $T_c$. When $T$ is
across $T_c$, the derivative of the QNS with respect to $T$ changes sign, which
is in agreement with the result in the present paper. However, when $T$ is
around $T_c$, the QNS calculated in Ref. \cite{He2} is larger than the free
quark gas result. This point is hard to understand physically, because the free quark
gas result should be an upper limit and a result beyond this limit means the
space-like damping mode is extremely dominant in the energy spectrum
of quark matter. According to the result of the present paper, such a
behavior might be an artifact of the model gluon propagator choosed in  Ref. \cite{He2}. In addition, from Fig. 5 it can also be seen that for both RL and BC trancation schemes, the bahavior of the QNS across the phase transition point is nonmonotonic. Such a nonmonotonic behavior of the QNS can be regarded as a prediction of our model.
Here it should be pointed out that the nonmonotonic behavior of QNS around $T_c$ obtained in our model differs from the existing lattice results of Refs. \cite{L5,L6} where a monotonic behavior of the QNS across the phase transition point is
found.  It should be noted that in the present work we are working in the chiral limit, whereas in the lattice simulations of Refs. \cite{L5,L6} a nonzero current quark mass is assumed. It is likely that the nonmonotonic behavior of QNS around $T_c$ in the case of chiral limit obtained in the present work is not within reach of the existing lattice studies.

\begin{figure}[t]
\begin{minipage}[t]{7cm}
\includegraphics[width=6.5cm]{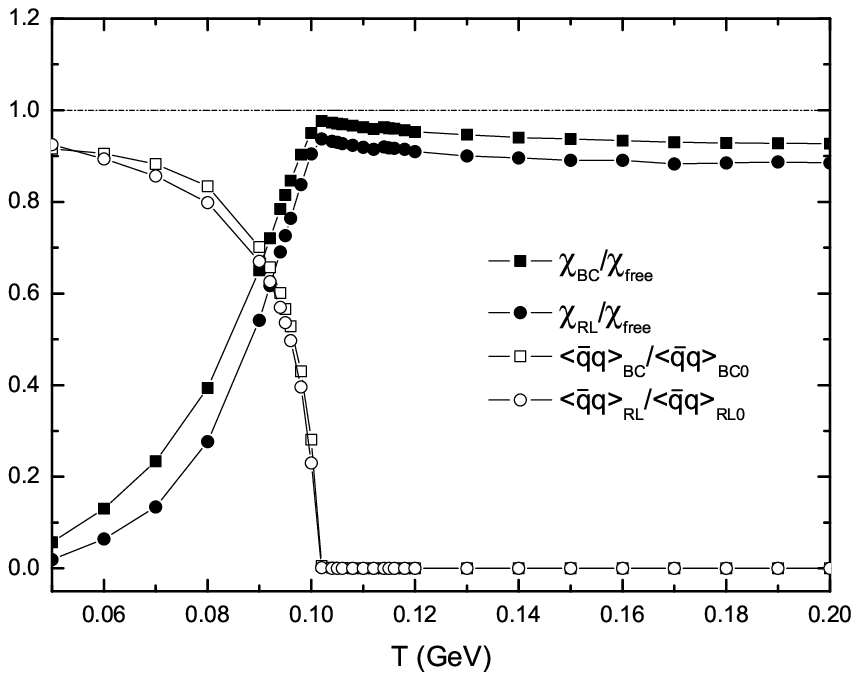}
\caption{Temperature dependence of the QNS and the chiral condensate.}
\end{minipage}
\hfill
\begin{minipage}[t]{7cm}
\includegraphics[width=6.5cm]{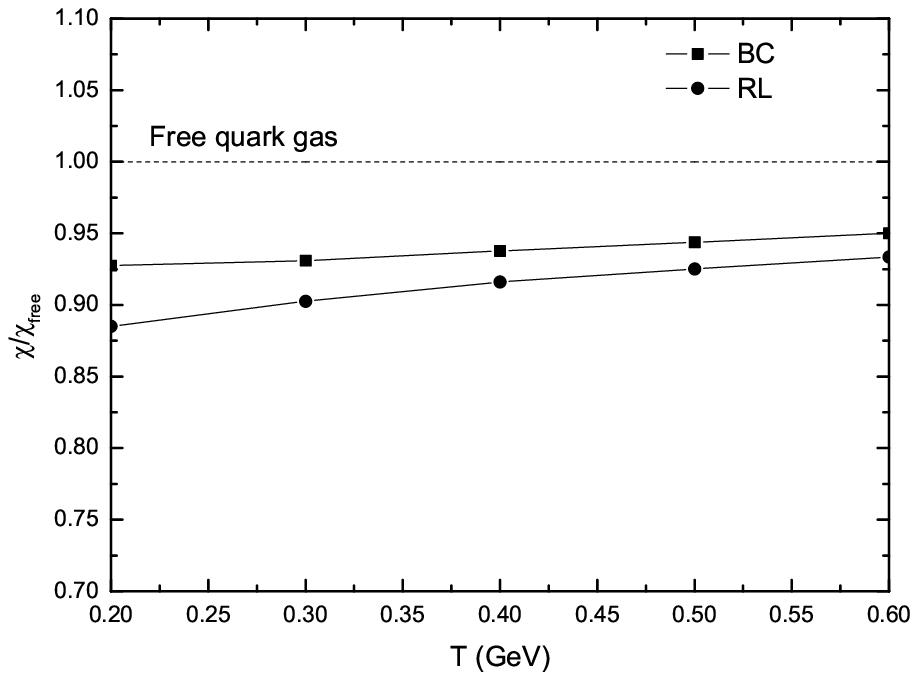}
\caption{The QNS in the range $T>200$ MeV }
\end{minipage}
\end{figure}

\section{Summary}

In summary, we calculate the QNS of QCD at finite temperature using both
RL and BC-type truncation schemes of DSEs approach. Our result shows
that when $T<T_c$, the QNS increases with increasing $T$ and the rate
of increasing also increases with $T$. At $T_c$ the derivative of
QNS with respect to $T$ undergoes a sudden change (from positive value to negative value). Such a
behavior can be regarded as the signal of chiral phase transition.
When $T_c<T\lesssim200$ MeV, the QNS decreases slowly with
increasing $T$. When $T$ is larger than about $200$ MeV, the QNS again
increases very slowly towards the free quark gas limit with
increasing $T$, which is the result of asymptotic freedom of QCD.
When $T\gtrsim95$ MeV, the difference between the QNS obtained using RL truncation and the one obtained using BC-type truncation is small ($<10\%$), which shows that the dressing effect of the quark-gluon vertex on the QNS is small.
In addition, from our study it is found that for both RL and BC truncation schemes, the bahavior of the QNS across the phase transition point is nonmonotonic. Such a nonmonotonic behavior of QNS around $T_c$ obtained in our model differs from the existing lattice results of Refs. \cite{L5,L6} where a monotonic behavior of the QNS across the phase transition point is
found. This difference deserves further study. Finally, we want to stress that the
method used in the present paper can be generalized to the case of both finite temperature and finite chemical potential and be used to investigate the properties of the critical end point and the phase diagram of QCD.

\acknowledgments

This work is supported in part by the Postdoctoral Science
Foundation of China (under Grant No. 20100471308), the National Science Foundation of China (under Grant Nos 11075075 and 10935001) and the Research Fund for the Doctoral Program of Higher Education (under Grant No 200802840009).

\appendix
\section{Expressions of the BC-type kernel}

From general Lorentz analysis one can find the vector vertex
$\Gamma_\mu(p_k,0)$ has the following structure ($i=1,2,3$)
\begin{eqnarray}
\Gamma_i(p_k,0)&=&E_0\gamma_i+p_i[E_1 \vec\gamma\cdot\vec
p+E_2\omega_k\gamma_4-iE_3]\\
\Gamma_4(p_k,0)&=&\tilde{F}_0\gamma_4+\omega_k
(F_1\vec\gamma\cdot\vec p+\tilde{F}_2\omega_k\gamma_4-iF_3)\\
&=&\omega_k(F_1\vec\gamma\cdot\vec p+F_2\omega_k\gamma_4-iF_3)
\end{eqnarray}
where $F_2\equiv\tilde{F}_2+\tilde{F}_0/\omega_k^2$,
$E_0,\ldots,E_3$ and $\tilde{F}_0,F_1,\tilde{F}_2,F_2,F_3$ are
scalar functions of $\vec p^2$ and $\omega_k^2$. For the free field case,
$E_0=\tilde{F}_0=1$, $E_1=E_2=E_3=F_1=\tilde{F}_2=F_3=0$ and hence
$F_2=1/\omega_k^2$.

From Eq. (\ref{vWTI}) the four-point Schwinger function
$\Lambda_{\mu\sigma}(p_k,q_n;0)$ can be written as follows
($i,j=1,2,3$)
\begin{eqnarray}
i\Lambda_{ij}(p_k,q_n;0)&=&\frac{(p_j+q_j)}{2}
\Big[2\Delta_{E_0}\gamma_i+\Delta_{E_1}(p_i\vec\gamma\cdot\vec
p+q_i\vec\gamma\cdot\vec q)+\Delta_{E_2}(p_i\omega_k+q_i\omega_n)
\gamma_4\nonumber\\
&&{}-i\Delta_{E_3}(p_i+q_i)\Big]+\Sigma_{E_1}\left[\frac{p_i+q_i}{2}
\gamma_j+\frac{\vec\gamma\cdot(\vec p+\vec
q)}{2}\delta_{ij}\right]\nonumber\\
&&{}+\Sigma_{E_2}\frac{\omega_k+\omega_n}
{2}\gamma_4\delta_{ij}-i\Sigma_{E_3}\delta_{ij}\\
i\Lambda_{i4}(p_k,q_n;0)&=&\frac{\omega_k+\omega_n}{2}\Big[2\Delta_{E_0}
\gamma_i+\Delta_{E_1} (p_i\vec\gamma\cdot\vec
p+q_i\vec\gamma\cdot\vec
q)+\Delta_{E_2}(p_i\omega_k+q_i\omega_n)\gamma_4\nonumber\\
&&{}-i\Delta_{E_3}(p_i+q_i)\Big]+\Sigma_{E_2}\frac{p_i+q_i}{2}\gamma_4\\
i\Lambda_{4i}(p_k,q_n;0)&=&\frac{p_i+q_i}{2}\Big[\Delta_{F_1}
(\omega_k\vec\gamma\cdot\vec p+\omega_n\vec\gamma\cdot\vec
q)+\Delta_{F_2}(\omega_k^2+\omega_n^2)\gamma_4\nonumber\\
&&{}-i\Delta_{F_3} (\omega_k+\omega_n)\Big]+\Sigma_{F_1}
\frac{\omega_k+\omega_n}{2}\gamma_i\\
i\Lambda_{44}(p_k,q_n;0)&=&\frac{\omega_k+\omega_n}{2}
\Big[\Delta_{F_1} (\omega_k\vec\gamma\cdot\vec
p+\omega_n\vec\gamma\cdot\vec
q)+\Delta_{F_2}(\omega_k^2+\omega_n^2)\gamma_4\nonumber\\
&&{}-i\Delta_{F_3} (\omega_k+\omega_n)\Big]+\Sigma_{F_1}
\frac{\vec\gamma\cdot(\vec p+\vec q)}{2}+\Sigma_{F_2}
(\omega_k+\omega_n)\gamma_4-i\Sigma_{F_3}
\end{eqnarray}

\end{document}